\documentclass{article}

\usepackage{arxiv}

\usepackage{physics}
\usepackage[utf8]{inputenc} % allow utf-8 input
\usepackage[T1]{fontenc}    % use 8-bit T1 fonts
\usepackage{hyperref}       % hyperlinks
\usepackage{url}            % simple URL typesetting
\usepackage{booktabs}       % professional-quality tables
\usepackage{amsfonts}       % blackboard math symbols
\usepackage{amsmath}        % for \text command
\usepackage{nicefrac}       % compact symbols for 1/2, etc.
\usepackage{microtype}      % microtypography
\usepackage{lipsum}		% Can be removed after putting your text content
\usepackage{graphicx}
\usepackage[numbers]{natbib} % Enables numbered/squared citations
\usepackage{doi}
\usepackage{caption}
\usepackage{subcaption}
\usepackage{algorithm}
\usepackage{algpseudocode}
\usepackage{setspace}
\usepackage{ragged2e}
\usepackage{cleveref}

\title{Bayesian Optimization in Chemical Compound Sub-Spaces using Low-Dimensional Molecular Descriptors}

\author{ \href{https://orcid.org/0009-0000-7792-885X}
{\includegraphics[scale=0.06]{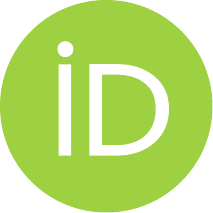}\hspace{1mm}Yun-Wen Mao}\\
	Department of Chemistry, University of British Columbia,\\
	Vancouver, B.C. V6T 1Z1, Canada \\
	Stewart Blusson Quantum Matter Institute,\\
	Vancouver, B.C. V6T 1Z4, Canada\\
	\texttt{ymaoai@student.ubc.ca} \\
	\And
	Roman V. Krems \\
	Department of Chemistry, University of British Columbia,\\
	Vancouver, B.C. V6T 1Z1, Canada \\
	Stewart Blusson Quantum Matter Institute,\\
	Vancouver, B.C. V6T 1Z4, Canada\\
	\texttt{rkrems@chem.ubc.ca} \\
}

\renewcommand{\shorttitle}{Bayesian optimization in chemical compound sub-spaces}

\hypersetup{
pdftitle={Bayesian Optimization in Chemical Compound Sub-Spaces using Low-Dimensional Molecular Descriptors},
pdfsubject={physics.chem-ph },
pdfauthor={Yun-Wen Mao, Roman V. Krems},
pdfkeywords={First keyword, Second keyword, More},
}

\begin{document}
\maketitle

\begin{abstract}
	Efficient optimization of molecules with targeted properties remains a significant challenge
	due to the vast size and discrete nature of chemical compound space.
	Conventional machine-learning-based optimization approaches typically require large datasets to construct accurate surrogate models, limiting their applicability in data-scarce settings.
	In this study, we present a Bayesian optimization (BO) framework
	that identifies optimal molecular
	structures with high precision using fewer than 2,000 training data points
	within a chemical subspace containing more than 133,000 molecules.
	The framework employs a low-dimensional and
	physics-informed molecular descriptor vector
	that facilitates data-efficient surrogate modeling and optimization.
	A key innovation of the proposed framework is a reliable inverse mapping scheme
	that translates optimized points in the descriptor space
	back into chemically valid molecular structures,
	thereby bridging continuous optimization and discrete molecular design.
	We demonstrate the effectiveness of our approach on the QM9 benchmark dataset,
	where the framework successfully identifies organic molecules with the target entropy and zero-point vibrational energy (ZPVE) values.
	For entropy optimization, our approach achieves a 100$\%$ success rate while requiring fewer than 1,000 molecular evaluations in more than 80$\%$ of test cases.
	For ZPVE, the success rate exceeds 80$\%$ for molecules containing more than two heavy atoms.
	These results highlight the critical role of low-dimensional, interpretable descriptors
	in enabling data-efficient optimization and robust inverse molecular design,
	and establish Bayesian optimization as a practical tool for molecular discovery in small-data regimes.
\end{abstract}

% keywords can be removed

\section{Introduction}\label{sec:intro}
The chemical compound space is vast due to the combinatorial nature of molecular structures.
The number of molecules suitable for medical applications is estimated to range from $10^{23}$ to $10^{60}$~\cite{polishchuk2013estimation}.
Combined with the high cost of experimental screening and quantum-chemistry calculations, the size of the chemical compound space makes the search for molecules with desired physical and chemical properties extremely challenging.
To address this challenge, data-driven approaches---
including machine learning (ML)--- have emerged as important tools. 
ML models can serve as surrogates for experiments or quantum-chemistry computations, enabling efficient screening of chemical compound sub-spaces, such as sets of structural isomers or specific classes of ligands.
Various ML methods have been employed for molecular property predictions,
including feed-forward neural networks (NNs)~\cite{sureyyarifaioglu_deepscreen_2020, st_john_message_passing_2019, gomez_bombarelli_design_2016},
large language models (LLMs)~\cite{caldasramos_llm_review_2025, bagal_molgpt_2022, wang_cmolgpt_2023, ye_novo_2024, frey_neural_2023, ye_drugassist_2025},
generative adversarial networks (GANs)~\cite{zeni_generative_2025, kim_generative_2020, zhao_high-throughput_2021, liu_generative_2023},
variational autoencoders (VAEs)~\cite{gomez-bombarelli_automatic_2018, ma_constrained_2018, jin_junction_2018, iwata_vgae-mcts_2023, fallani_inverse_2024},
kernel methods~\cite{duvenaud2013structure, khan2023quantum, Mao_2024, bigi_wigner_2024},
random forests~\cite{popova_deep_2018}, 
and reinforcement learning (RL)~\cite{zhou_optimization_2019}.
These approaches have demonstrated remarkable success in accelerating molecular discovery. 
However, high-throughput screening often trades accuracy for speed,
and may therefore overlook important molecules.
An alternative strategy is direct optimization within chemical compound sub-spaces to identify specific molecules with high precision.
High-precision optimization of molecular properties should be validated using experiments or quantum-chemical calculations to confirm the candidates proposed by the optimizer.
This requires efficient optimization schemes,
capable of converging to optimal solutions in complex molecule spaces with a small number of confirmations.
The development of such efficient optimizers faces two challenges: (i) molecules are complex and generally require high-dimensional numerical descriptors, which leads to high-dimensional optimization problems;
and (ii) optimization requires the inverse mapping from numerical descriptors to molecules. 

A great variety of molecular descriptors have been developed to 
encode compositional, structural, and electronic property information
into numerical features suitable for learning algorithms.
It has been established that physics-based descriptors generally
yield more data-efficient models than, for example, descriptors derived from tokenized SMILES strings processed as text~\cite{musil_physics_2021, wigh_review_2022, ding_exploring_2024, meng_when_2025}.
Examples of molecular descriptors that incorporate physical features include the Behler-Parrinello symmetry functions (ACSF)~\cite{behler_ACSF_2011, Gastegger2018}, smooth overlap of atomic positions (SOAP)~\cite{SOAP_2012}, local many-body tensor representation (LMBTR) functions~\cite{huo_unified_2022}, Faber-Christensen-Huang-Lilienfeld (FCHL19)~\cite{FCHL19}, convolutional many-body distribution functionals (cMBDF)~\cite{khan_generalized_2025}, Coulomb matrices~\cite{PRL_CM_2011}, bag of bonds (BoB) features~\cite{BoB_2015}, and spectrum of London and Axilrod-Teller-Muto (SLATM)~\cite{huang_quantum_2020}.
While these descriptors have enabled accurate ML predictions for a wide range of molecular properties, optimization within the high-dimensional chemical compound space represented by such descriptors remains highly challenging.
This difficulty arises from the fundamentally non-smooth nature of the objective landscape,
as molecules are inherently discrete graphs of atoms and bonds.
Thus, even small structural modifications, such as functional group substitutions,
can lead to abrupt changes in molecular properties~\cite{xia_understanding_2023}.
One way to circumvent this problem is to construct smooth and continuous
latent spaces using models such as VAEs~\cite{gomez-bombarelli_automatic_2018, ma_constrained_2018, jin_junction_2018, iwata_vgae-mcts_2023, fallani_inverse_2024},
GANs~\cite{Guimaraes_ORGAN_2018, decao_MolGAN_2022},
or diffusion models~\cite{liu_diffmeta_rl_2025, sako_diffint_2025, jin_liganddiff_2024},
enabling gradient-based optimization.
However, training such generative models typically requires large datasets.
Therefore, data-efficient and gradient-free methods are needed to allow direct optimization in chemical sub-spaces,
especially when data are scarce.

A possible approach to reduce the data requirements for molecular property discovery is Bayesian optimization (BO). 
BO is particularly well suited for expensive evaluations (i.e. experimental or quantum-chemical confirmation of molecular properties), gradient-free settings, and global optimization.
In chemistry and materials science, BO has been applied to diverse optimization problems,
including catalyst discovery~\cite{manoj_multi_objective_2023, wu_race_2024, tadepalli_smart_2025, feng_bayesian_2023},
molecular docking~\cite{bespalov_physics_informed_2025, cao_bayesian_2020},
and material property optimization~\cite{hase_gryffin_2021, kurunczi_papp_bayesian_2024, li_sequential_2024, pollice_data_driven_2021}.
However, Bayesian optimization is particularly sensitive to the dimensionality problem and,
as other optimization approaches, requires overcoming the inverse problem.
High dimensionality of the input space limits the probabilistic surrogate models underlying BO.  
Although probabilistic models provide uncertainty estimates that guide the search, 
their predictive performance deteriorates rapidly in
high-dimensional spaces~\cite{Chen02092023}.
This issue is especially pronounced in chemistry,
where high-quality data are scarce,
yet molecular descriptor spaces are often high-dimensional
in order to capture subtle structural variations
and complex chemical features.
Prior work~\cite{Willatt_2018} has shown that
reducing descriptor dimensionality can significantly improve both
ML model accuracy and computational efficiency.
Furthermore, other studies have demonstrated that
physically grounded descriptors
result in more robust and generalizable models~\cite{musil_physics_2021, bigi_wigner_2024}.
Our previous work~\cite{Mao_2024}
introduced physically motivated and
low-dimensional molecular descriptor vectors that preserve essential
chemical information while substantially reducing dimensionality.
This representation enables accurate Gaussian process regression (GPR) interpolation
even with limited data and provides the foundation for the optimization framework developed here.
In the present work, we exploit these low-dimensional descriptors to explore the feasibility of BO in chemical compound spaces. 
To achieve this, we develop and demonstrate an inverse-mapping scheme that converts molecular descriptor vectors back into molecules. 

Inverse design requires mapping points in descriptor space back to chemically valid molecular structures,
a key step in any molecular property optimization strategy.
This inverse mapping is difficult due to the discrete and constrained nature of molecules.
Most points in descriptor-property space do not correspond to physically-realizable molecules,
often making reconstruction from a set of optimized descriptors ill-posed.
Several strategies~\cite{sanchez_inverse_2018} have been proposed to address this challenge.
For high-dimensional descriptors, prior work explored
various deep learning techniques, including reinforcement learning (RL)~\cite{you_graph_2018, zhou_optimization_2019},
evolutionary approaches such as genetic algorithms (GA)~\cite{jensen_graph-based_2019, reeves_assessing_2020, henault_chemical_2020},
and generative models, VAEs~\cite{gomez-bombarelli_automatic_2018, ma_constrained_2018, jin_junction_2018, iwata_vgae-mcts_2023, fallani_inverse_2024},
generative adversarial networks (GANs)~\cite{Guimaraes_ORGAN_2018, decao_MolGAN_2022},
and generative pre-trained transformer (GPT) models~\cite{bagal_molgpt_2022, wang_cmolgpt_2023, ye_novo_2024, frey_neural_2023, ye_drugassist_2025}.
These models enable inverse molecular design
by generating molecules with targeted properties.
However, such approaches typically require large training datasets,
which are unavailable for many practical applications in chemistry.
Moreover, these methods often lack interpretability or chemical consistency,
especially when applied to small-data or physics-informed settings.
In addition, the latent representations learned by these models do not always align with
well-defined sets of molecular descriptors,
limiting integration with descriptor-based optimization frameworks.

In this work, we address both the dimensionality challenge and the inverse
problems inherent to optimization in chemical compound spaces.
Our approach introduces a general strategy for solving the inverse problem in low-dimensional,
physically motivated descriptor spaces, providing a reliable bridge between optimized descriptors and valid molecular structures.
Molecules are first represented using the compact and
physically inspired descriptors introduced in our previous work~\cite{Mao_2024}.
BO is then employed to efficiently explore this reduced descriptor space
using GPR with kernels optimized for data-efficient interpolation.
Each candidate vector of descriptors proposed by BO is mapped onto a molecule using an algorithm that
identifies chemically plausible structures by matching
the predicted stoichiometry and shape characteristics against a molecular database.
We validate this approach using the QM9 dataset,
demonstrating that the optimization scheme can successfully identify
organic molecules with desired entropy
and zero-point vibrational energy (ZPVE),
achieving up to 100$\%$ success rate with a small number of iterations. 
This framework enables the optimizer to propose candidates
in continuous descriptor spaces and realize them as valid molecules,
effectively bridging the gap between continuous optimization and discrete chemical design.

\section{Dataset summary}\label{sec:dataset}
We use the QM9 dataset~\cite{qm9_dataset} to benchmark our optimization framework.
The dataset contains 133,885 stable organic molecules composed of the following atoms: H, C, O, N, and F.
Within the QM9 dataset, there are 526 subsets of constitutional isomers defined as molecules with the same chemical formula but different atom connectivity. 
The subsets range in size from two molecules to as many as 6,059 molecules, with the largest subset corresponding to the chemical formula ${\rm C_{7}H_{10}O_{2}}$.
For each molecule in QM9,
properties such as zero-point vibrational energy (ZPVE), enthalpy, and Gibbs free energy are computed at the B3LYP/6-31G(2df,p) level of theory.
In this work,
we focus on optimizing the molecular entropy and ZPVE,
as our previously developed low-dimensional descriptors~\cite{Mao_2024}
are particularly well-suited for interpolating these properties,
yielding interpolation models with chemical accuracy $ \leq 1~{\rm kcal\,mol^{-1}}$
using a small number of molecules in the training set.

\section{Method}
%%%%%%%%%%%%%%%%%%%%%%%%%%%%%%%%%%%%%%%%%%%%%%%%%%%%%%%%%%%%%%%%%%%%%%%%%%%%%%%%%%%%%%%%%%%%%%%%%%%%%%%%%%%%%%%%%%%%%%%%%%%%%%%%%%%%%%%%
\subsection{Molecular property optimization scheme}\label{subsec:mol_opt}
The pseudo-algorithm~\ref{algo:molecule_opt} outlines the iterative process of the present molecular property optimization framework.
The goal of the algorithm is to identify the optimal molecular structure with the property value closest to a specified target value $y_\text{target}$ within a given chemical sub-space.
The objective function is defined as
\begin{equation}\label{eq:delta_i}
    \delta(x)= |y_{\rm target}-\gamma(x)|, 
\end{equation}
where $x$ is a vector of numerical features representing a molecule,
and $\gamma(\cdot)$ denotes the underlying ML model that predicts the molecular property (e.g., a GP model).
The optimization seeks the global minimum of $\delta(x)$,
\begin{equation}\label{eq:BO_objective_function}
    x^* = \arg \min_{x} \delta(x),
\end{equation}
where $x^*$ is the vector of features corresponding to the optimal molecule. 
Training the optimization algorithm requires data points sampled across the input space.
The training dataset with $N_\text{train}$ samples is defined as
$\mathcal{D}_\text{train} = (X_\text{train}, \Delta_\text{train})$,
where the $d$-dimensional inputs are
$X_\text{train}=\{x_i\mid i\in [1, N_\text{train}]\} \subset \mathbb{R}^{d}$
and the corresponding outputs are
$\Delta_\text{train} =\{\delta(x_i) \mid i\in [1, N_\text{train}]\}$.
%%%%%%%%%%%%%%%%%%%%%%%%%%%%%%%%%%%%%%%%%%%%%%%%%%%%%%%%%%%%%%%%%%%
\begin{algorithm}\captionsetup{justification=justified}
\setstretch{1.2} 
\caption{\justifying Bayesian Optimization with Chemical Formula Search}\label{algo:molecule_opt}
\begin{algorithmic}[1]
\State Initialize training data using Latin hypercube sampling $X_{\rm train}=\{x_i\mid i\in [1, N_{t=0}]\} \subset \mathbb{R}^{d}$
% \State Compute $\delta_i =\texttt{InverMap}(x_i)$ for each $x_i \in X_{\rm train}$ to form $\mathcal{D}_{\rm train} = (X_{\rm train}, \Delta_{\rm train})$
\State Initialize output set $\Delta_\text{train} = \emptyset$
\For{$i = 1$ to $N_0$}
    \State Compute $\delta_i = \delta(x_i)$ for $x_i \in X_\text{train}$
    \State Update $\Delta_\text{train} = \Delta_\text{train} \cup \{\delta_i\}$
\EndFor
\State Form the initial training dataset $\mathcal{D}_\text{train} = (X_\text{train}, \Delta_\text{train})$
\While{$\min(\Delta_{\text{train}}) > \epsilon$}
    \State Use Bayesian optimization to suggest next point: $x_{\text{BO}} = \texttt{BayesianOptimization}(\mathcal{D}_{\rm train})$
    \State Compute $(x_t, \delta_t)$, where $\delta_t =\delta(x_{\text{BO}})$, for the $t^{\rm th}$ BO iteration\label{algo_line:inverse_map}
    \State Update $\mathcal{D}_{\text{train}} = \mathcal{D}_{\text{train}} \cup \{x_t, \delta_t\}$
\EndWhile
\State \Return $x_t$
\end{algorithmic}
\end{algorithm}
%%%%%%%%%%%%%%%%%%%%%%%%%%%%%%%%%%%%%%%%%%%%%%%%%%%%%%%%%%%%%%%%%%%
The initial training dataset $X_{\rm train}$ is generated using Latin-hypercube sampling (LHS)~\cite{LHS_Mckay, LHS_Stein},
a statistical randomized sampling method that prevents clustering in the input space.
Each input point $x_i\in X_\text{train}$ is mapped to an output value $\delta_i$ using the inverse mapping scheme described in Sec.~\ref{subsec:MolMap}.
Given the training dataset $\mathcal{D}_\text{train}$,
Bayesian optimization proposes a new sampling point $x_{\rm BO}$.
Using the mapping function mentioned in line 4 of Algorithm~\ref{algo:molecule_opt},
this vector is mapped on the corresponding molecule and hence the output property value $\delta_\text{BO}$ is computed.
The training set $\mathcal{D}_\text{train}$ is updated to include $
(x_\text{BO}, \delta_\text{BO})$ and the process is iterated.
The iterative optimization process terminates when $\min(\Delta_\text{train})<\epsilon$, where $\epsilon$ is a predefined error threshold.
A nonzero threshold $\epsilon > 0$ permits limited error tolerance, accounting for thermal noise and model limitations.
For example, $\epsilon = 1$ ${\rm kcal\,mol^{-1}}$ corresponds to chemical accuracy, which is defined as the maximum acceptable deviation between theoretical predictions and experimental measurements due to thermal fluctuations~\cite{RevModPhys.71.1267}. 
%%%%%%%%%%%%%%%%%%%%%%%%%%%%%%%%%%%%%%%%%%%%%%%%%%%%%%%%%%%%%%%%%%%%%%%%%%%%%%%%%%%%%%%%%%%%%%%%%%%%%%%%%%%%%%%%%%%%%%%%%%%%%%%%%%%%%%%%
\subsection{Details of Bayesian Optimization}\label{subsec:BO}
Bayesian optimization (BO) is an iterative ML algorithm designed to efficiently locate the global minimum of an objective function.
In this work, we use the objective function defined in~\cref{eq:BO_objective_function}.
BO relies on probabilistic surrogate models trained on the dataset $\mathcal{D}_\text{train}$,
which provide both predictions of the black-box function and associated uncertainty estimates.
At each iteration, BO selects a sampling point by maximizing an acquisition function $\alpha(x)$, which leverages the output of the surrogate model to balance exploration and exploitation.
Because the acquisition function is computationally inexpensive relative to experiments or quantum chemistry calculations,
it efficiently guides the search towards an optimal molecular structure.

%%%%%%%%%%%%%%%%%%%%%%%%%%%%%%%%%%%%%%%%%%%%%%%%%%%%%%%%%%%%%%%%%%%%%%%%%%%%%%%%%%%%%%%%%%%%%%%%%%%%%%%%%%%%%%%%%%%%%%%%%%%%%%%%%%%%%%%%
We use Gaussian Process regression (GPR)~\cite{murphy2018machine} as the surrogate model for BO.
GPR is a probabilistic, non-parametric supervised learning method that models the objective function $d(x)$ as a random function with a joint Gaussian prior over the function values. 
Given an unseen input set $X_*$, GPR predicts the corresponding outputs $\Delta_*$ by modeling the joint distribution between the training outputs $\Delta$ and the test output $\Delta_*$ as a multivariate Gaussian 
%%%%%%%%%%%%%%%%%%%%%%%%%%%%%%%%%%%%%%%%%%%%%%%%%%%%%%%%%%%%%%%%%%%%%%%%%%%%%%%%%%%%%%%%%%%%%%%%%%%%%%%%%%%%%%%%%%%%%%%%%%%%%%%%%%%%%%%%
\begin{equation}\label{gp_definition}
    \begin{pmatrix}
    \Delta\\
    \Delta_{*}
    \end{pmatrix}\sim N\left(\begin{pmatrix}
    \mu\\
    \mu_*
    \end{pmatrix},\begin{pmatrix}
    K & K_* \\
    K_*^{T} & K_{**}
    \end{pmatrix}\right)
\end{equation}
%%%%%%%%%%%%%%%%%%%%%%%%%%%%%%%%%%%%%%%%%%%%%%%%%%%%%%%%%%%%%%%%%%%%%%%%%%%%%%%%%%%%%%%%%%%%%%%%%%%%%%%%%%%%%%%%%%%%%%%%%%%%%%%%%%%%%%%%
where $\mu = [m(x_1), \dots, m(x_{N_{\rm test}})]$ and $\mu_* = [m(x_{*,1}), \dots, m(x_{*,N_{\rm train}})]$ denote the prior mean functions evaluated at the training and test points, respectively, with $m(\cdot)$ representing the prior mean. The covariance matrices are defined using the kernel function $k$, such that $K = k(X_t, X_t)$, $K_*=k (X_t,X_*)$, and $K_{**}=k(X_*,X_*)$. 
As discussed in our previous work~\cite{Mao_2024}, kernel optimization is critical for accurate interpolation. 
We use the Bayesian information criterion (BIC)~\cite{duvenaud2011additive, duvenaud2013structure} as the model selection metric to construct kernel functions of variable complexity by forming linear combinations and products of basis kernel functions.
The basis kernel functions used in this work are rational quadratic kernel ($\kappa_{\rm RQ}$)
\begin{equation}
    \kappa_{\rm RQ} (x_i, x_j) = \left(1+\frac{d(x_i, x_j)^2}{2a l^2}\right)^2
\end{equation}
where $a$ is the scale mixture parameter, $l$ is the length scale of the kernel, and $d(\cdot,\cdot)$ is the Euclidean distance between two data points;
Mat\'{e}rn kernel($\kappa_{\rm MAT}$)
\begin{equation}
    \kappa_{\rm MAT} (x_i, x_j) = \frac{1}{\Gamma(\eta)2^{\eta-1}}\left(\frac{\sqrt{2\eta}}{l}d(x_i, x_j)\right)^\eta K_{\eta}\left(\frac{\sqrt{2\eta}}{l}d(x_i, x_j)\right)
\end{equation}
where $\eta$ and $l$ are positive parameters,
$K_{\nu}(\cdot)$ is a modified Bessel function,
and $\Gamma (\cdot)$ is the gamma function;
and dot product kernels ($\kappa_{\rm DP}$)
\begin{equation}
    \kappa_{\rm DP}(x_i, x_j) = \sigma_0^2 + \braket{x_i, x_j}
\end{equation}
where $\sigma^{2}_{0}$ is a parameter which controls the inhomogeneity of the kernel.
% {\color{red} Give the explicit equations for the basis kernel functions}

The acquisition function is constructed from the predictive mean $\mu_*$ and uncertainty $\sigma_*$ provided by the GPR model.
For an unseen input point $x_*$, the predictive mean and variance of a noiseless GPR model are given by
\begin{align}
    \mu(x_*) = k^{\rm T}_*K^{-1}\Delta, \\
    \sigma_* = k(x_*, x_*)-k*^{\rm T}K^{-1}k_*,
\end{align}
where $k_{*}$ is a vector with entries $k(x_*, x_i)$
for $i\in[1, N_\text{train}]$.
These quantities are used to compute the acquisition function.
Various forms of $\alpha(x)$ have been proposed~\cite{wang2023recent},
each balancing exploration and exploitation differently. 
Here, we use the upper confidence bound (UCB) acquisition function 
defined as
\begin{equation}\label{alpha_ucb}
    \alpha(x) = \mu_* (x) + \zeta \sigma_*(x),
\end{equation}
where the tunable parameter $\zeta$ controls the trade-off between exploration, driven by the uncertainty term $\sigma_*(x)$, and exploitation, guided by $\mu_*(x)$, across the space of molecules represented by multi-dimensional descriptor vectors. 
%%%%%%%%%%%%%%%%%%%%%%%%%%%%%%%%%%%%%%%%%%%%%%%%%%%%%%%%%%%%%%%%%%%%%%%%%%%%%%%%%%%%%%%%%%%%%%%%%%%%%%%%%%%%%%%%%%%%%%%%%%%%%%%%%%%%%%%%
\subsection{Molecular descriptors}\label{subsec:molecular_descriptor}
Computational and machine learning models require  molecules to be represented by numerical vectors.
In general, molecular descriptors are numerical features or binary strings constructed through mathematical transformations that encode structural properties of molecules.
Many examples of physics-inspired molecular descriptors have been proposed, where these descriptors can be broadly categorized as either `local' or `global' descriptors.
Local descriptors represent the atomic environment within a molecule.
Examples include smooth overlap of atomic positions (SOAP)~\cite{SOAP_2012}, Faber-Christensen-Huang-Lilienfeld (FCHL19)~\cite{FCHL19}, Behler-Parrinello symmetry functions (ACSF)~\cite{behler_ACSF_2011, bespalov_physics_informed_2025}, and LMBTR~\cite{huo_unified_2022}.
Global descriptors characterize the molecule as a whole. Descriptors such as Coulomb matrices~\cite{PRL_CM_2011}, bag of bonds (BoB) features~\cite{BoB_2015}, and spectrum of London and Axilrod-Teller-Muto (SLATM)~\cite{huang_quantum_2020} are under this category.
Combined with optimized learning algorithms, these representations allow ML models to interpolate chemical sub-spaces efficiently. 

In this work, we use the physically motivated, low-dimensional set of molecular descriptors introduced in our previous work~\cite{Mao_2024}.
We begin by forming three-dimensional vectors derived from the Coulomb matrix (CM).
The resulting numerical representation can be viewed as a predominantly global descriptor.
Each element $M_{ij}$ in CM is defined as
\begin{equation}\label{eq:coulomb_mat}
  M_{ij} =
    \begin{cases}
      0.5Z_{i}^{2.4} & \text{for $i=j$}\\
      \frac{Z_i Z_j}{R_{ij}} & \text{for $i\neq j$},
    \end{cases}      
\end{equation}
where $Z_i$ and $Z_j$ are the atomic numbers of atoms $i$ and $j$, respectively, and $R_{ij}$ is the separation between atoms $i$ and $j$.
The matrix $M$ is reduced to a three-dimensional eigenvalue descriptor vector
\begin{equation}\label{eq:lambda_descriptor}
    \Lambda = \left[\lambda_\text{max}, \mu(\lambda), \sigma(\lambda)\right],
\end{equation}
where $\lambda_{\rm max}$ is the largest eigenvalue of $M$, and $\mu(\lambda)$ and $\sigma(\lambda)$ are the mean and standard deviation of the distribution of eigenvalues of $M$.
This vector provides a compact global characterization of the molecule.
%%%%%%%%%%%%%%%%%%%%%%%%%%%%%%%%%%%%%%%%%%%%%%%%%%%%%%%%%%%%%%%%%%%%%%%%%%%%%%%%%%%%%%%%%%%%%%%%%%%%%%%%%%%%%%%%%%%%%%%%%%%%%%%%%%%%%%%%%%

The three-dimensional vectors (\ref{eq:lambda_descriptor}) are supplemented by inner products $\langle f_{Z}, f_m\rangle$, introduced and described in detail 
 in our previous work~\cite{Mao_2024}. Briefly, $\langle f_{Z}, f_m\rangle$ is defined as the inner product of an atomic reference probability density function $f_{Z}$ with the molecular function $f_{m}$.
The atomic reference distribution $f_{Z}$ of an atom with the nuclear charge $Z$ is
\begin{equation}\label{eq:N_Z}
    f_{Z} = \omega_{Z} \times f(x \mid \mu = (0.5 \times Z^{2.4})^{\beta}, \sigma^2 = \sigma^2_{Z}),
\end{equation}
where $\omega_{Z}$ is a weight constant, $\beta$ is a hyperparameter controlling the mean position, and $\sigma^2_{Z}$ sets the variance of the normal distribution $f(\cdot)$.
The definition of the molecular distribution $f_{m}$ is inspired by the Gershgorin circle theorem \cite{gershgorin1931uber, saad2011numerical_gershgorin}. We define $f_m$
as the sum of normal distributions $f(\cdot)$
\begin{equation}\label{eq:f_molecule}
    f_{m} = \sum_{i}^{n_\text{max}}\omega \times  f \left( x \mid \mu = (0.5\times Z_{i}^{2.4})^{\beta}, \sigma^2 = \left(\sum^{n_\text{max}}_{j\neq i}\frac{Z_{i} Z_{j}}{R_{ij}}\right)^s \right),
\end{equation}
where $n_\text{max}$ is the maximum number of atoms in molecule $m$,
$\omega$ is a weight constant, $\beta$ is a global hyperparameter shared with~\cref{eq:N_Z},
and $s$ adjusts the standard deviation. 
The constants and hyperparameters in~\cref{eq:N_Z}
and~\cref{eq:f_molecule} are adjustable.
%Further details can be found in Ref.~\cite{Mao_2024}.
While $\langle f_{Z}, f_m\rangle$ is fundamentally a global descriptor, since $f_{Z}$ targets only atoms with the nuclear charge $Z$,
it also encodes information about local atomic environments through the $R_{ij}$-dependent variance.

\subsection{Inverse mapping: from  descriptor space to molecules}\label{subsec:MolMap}
When a Bayesian optimizer proposes a sampling point in the descriptor space,
the corresponding numerical vector must be mapped onto a molecule,
as indicated in~\cref{algo_line:inverse_map} of~\cref{algo:molecule_opt}.
However, due to the discrete and constrained nature of molecules, mapping a descriptor vector back to a chemically valid molecular structure is challenging.
To address this challenge,
we develop Algorithm~\ref{algo:des_map_delta}.
Given a descriptor vector $x'_t$, the algorithm predicts the chemical formula ${\rm C}_{\nu_{\rm C}}{\rm H}_{\nu_{\rm H}}{\rm N}_{\nu_{\rm N}}{\rm O}_{\nu_{\rm O}}{\rm F}_{\nu_{\rm F}}$, where $\nu_{\rm atom}$ denotes the number of atoms in a molecular species. 
The resulting molecular formula can then be matched to species in a molecular database. 
In this work, we use the QM9 database as the reference; however, the procedure is general and can be applied to any molecular database or modules that generate valid molecular structures.
%%%%%%%%%%%%%%%%%%%%%%%%%%%%%%%%%%%%%%%%%%%%%%%%%%%%%%%%%%%%%%%%%%%%%%%%%%%%%%%%%%%%%%%%%%%%%%%%%%%%%%%%%%%%%%%%%%%%%%%%%%%%%%%%%%%%%%%%%%
% \begin{figure}[!h]
%     \centering
%     \captionsetup{justification=justified}
%     \includegraphics[width=0.95\linewidth]{figures/FIG2/figure.pdf}
%     \caption{\justifying Schematic diagram of the algorithm mapping a point in the descriptor space to a molecular structure.}
%     \label{fig:fig2}
% % \end{figure}
%%%%%%%%%%%%%%%%%%%%%%%%%%%%%%%%%%%%%%%%%%%%%%%%%%%%%%%%%%%%%%%%%%%%%%%%%%%%%%%%%%%%%%%%%%%%%%%%%%%%%%%%%%%%%%%%%%%%%%%%%%%%%%%%%%%%%%%%%%
%%%%%%%%%%%%%%%%%%%%%%%%%%%%%%%%%%%%%%%%%%%%%%%%%%%%%%%%%%%%%%%%%%%%%%
\begin{algorithm}\captionsetup{justification=justified}
\setstretch{1.2} 
\caption{\justifying Mapping from descriptor $x'_t$ to $(x_t, \delta_t)$}\label{algo:des_map_delta}
    \begin{algorithmic}[1]
    \State \label{line:ChemicalFormula}Generate the chemical formula from $x'_{t}$: ${\rm H}_{\nu_{\rm H}}{\rm C}_{\nu_{\rm C}}{\rm N}_{\nu_{\rm N}}{\rm O}_{\nu_{\rm O}}{\rm F}_{\nu_{\rm F}} = \texttt{ChemicalFormula}(x'_{t})$
    \State Search molecules with the correct stoichiometry: $\mathcal{M} = \texttt{DatabaseSearch}({\rm H}_{\nu_{\rm H}}{\rm C}_{\nu_{\rm C}}{\rm N}_{\nu_{\rm N}}{\rm O}_{\nu_{\rm O}}{\rm F}_{\nu_{\rm F}})$
    \If{$\mathcal{M} = \emptyset$}
        \State Set $x_t = x'_{t}$, $\delta_t = \delta_{\rm max}$
    \Else
        \State Find the best molecule: $m^{*} = \arg \min_{m\in \mathcal{M}}(\norm{\Lambda'_{i} - \Lambda_{m}}_2)$
    \EndIf
    \State Calculate descriptor: $x_t = \texttt{CalculateDescriptor}(m^{*})$ 
    \State Calculate the output: $\delta_t = d(x_t)$
    \State \Return $(x_t, \delta_t)$
    \end{algorithmic}
\end{algorithm}
%%%%%%%%%%%%%%%%%%%%%%%%%%%%%%%%%%%%%%%%%%%%%%%%%%%%%%%%%%%%%%%%%%%%%%%%%%%%%%%%%%%%%%%%%%%%%%%%%%%%%%%%%%%%%%%%%%%%%%%%%%%%%%%%%%%%%%%%%%
If no structure with the predicted formula exists in the database, the formula is considered chemically implausible.
The algorithm returns the original descriptor vector $x'_{t}$ with a penalty output value $\delta_{\rm max}$, providing negative feedback to the Bayesian optimization algorithm and discouraging exploration of chemically infeasible regions in the descriptor space.
The value of $\delta_{\rm max}$ is tunable and dataset-dependent.
In practice, we set $\delta_{\rm max} = 40\,{\rm kcal\,mol^{-1}}$ for molecular entropy optimization and $\delta_{\rm max} = 150\,{\rm kcal\,mol^{-1}}$ for zero-point vibrational energy (ZPVE) optimization, corresponding to the range of values of these properties in the QM9 dataset.

If valid candidates are found, the algorithm outputs a set of eligible molecular structures $\mathcal{M}$.
For each candidate $m\in\mathcal{M}$, the reduced three-dimensional Coulomb matrix eigenvalue descriptors $\Lambda$ are computed using~\cref{eq:lambda_descriptor}.
The most representative structure is selected from the set $\mathcal{M}$ according to
\begin{equation}\label{eq:m_star_selection}
    m^{*} = \arg \min_{m\in \mathcal{M}}(\norm{\Lambda'_{t} - \Lambda_{m}}_2),
\end{equation}
where $\Lambda'_{t}$ is the CM eigenvalue vector corresponding to the descriptor vector $x'_{t}$.
Practically, to efficiently identify $m^{*}$ within the large database, we use a kd-tree search~\cite{maneewongvatana1999analysis} implemented in the \texttt{Scipy} Python package.
As demonstrated in our previous work~\cite{Mao_2024}, the CM eigenvalue descriptor $\Lambda$ effectively encodes molecular shape information.
Thus,~\cref{eq:m_star_selection} ensures that the algorithm selects a molecule $m^*$ with the best matching structure among the constitutional isomers in $\mathcal{M}$. 
The corresponding output value $\delta_t$ is then computed using~\cref{eq:delta_i} and the algorithm returns the new point $(x_t, \delta_t)$.

\begin{figure}[!ht]
    \centering
    \captionsetup{justification=justified}
    \includegraphics[width=0.9\linewidth]{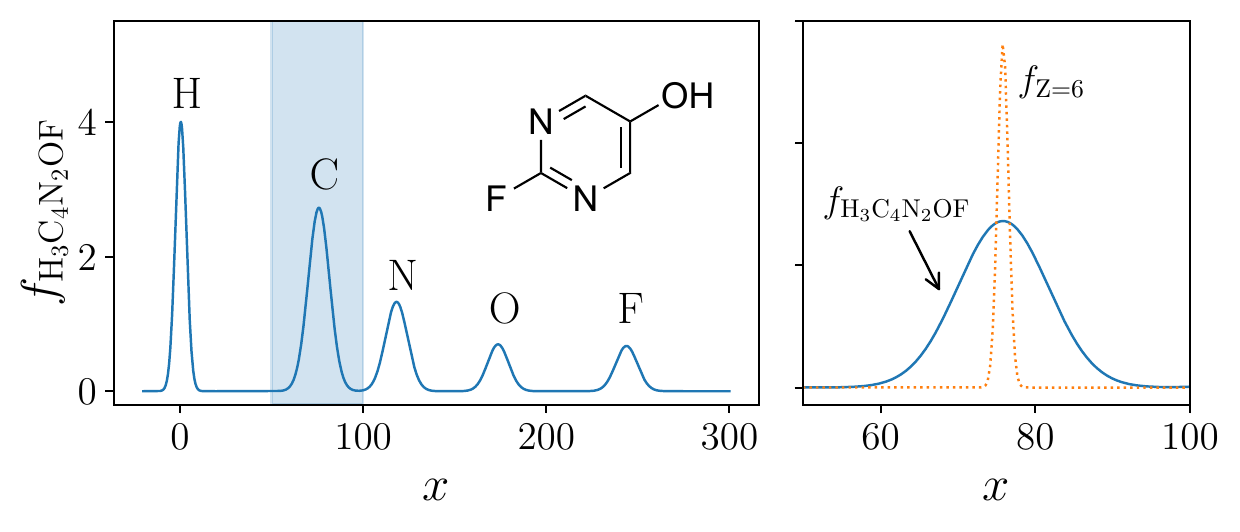}
    \caption{\justifying Left: Molecular function of ${\rm H_{3}C_{4}N_{2}OF}$ (blue solid line) computed from~\cref{eq:f_molecule}, with the shaded region highlighting the carbon peak.
    Right: Enhanced view of the carbon peak of
    $f_{\rm H_{3}C_{4}N_{2}OF}$ overlaid with $f_{Z=6}$
    (orange dotted line given by~\cref{eq:N_Z}).
    Parameters: $\omega_{Z} =10$, $\sigma_{Z}^{2} = 0.5$,
    $\beta = 1.2$, $s = 0.7$ and $\omega = 10$.}
    \label{fig:fig2}
\end{figure}

Given a molecule $m$, the classification scheme in~\cref{line:ChemicalFormula} of Algorithm~\ref{algo:des_map_delta}
determines the stoichiometric coefficients using
the set of molecular descriptors $\braket{f_{Z}, f_{m}}$
detailed in~\cref{subsec:molecular_descriptor}.
\cref{fig:fig2} illustrates how these descriptors encode stoichiometric information by comparing $f_{Z}$ and $f_{m}$ for the molecule ${\rm H_{3}C_{4}N_{2}OF}$.
The left panel of~\cref{fig:fig2} reveals five distinct peaks in $f_{m =\rm H_{3}C_{4}N_{2}OF}$, each centered at $x = 0.5Z^{2.4}$, corresponding to atoms of nuclear charge $Z$.
For example, the peak in the blue shaded region arises from the carbon atoms in ${\rm H_{3}C_{4}N_{2}OF}$, with the center located at $x = 0.5\cdot(6)^{2.4} = 75$. 
The right panel shows the overlap between the carbon peak and the carbon-atom reference probability density function, $f_{Z=6}$~\cref{eq:N_Z}.
%%%%%%%%%%%%%%%%%%%%%%%%%%%%%%%%%%%%%%%%%%%%%%%%%%%%%%%%%%%%%%%%%%%%%%%%%%%%%%%%%%%%%%%%%%%%%%%%%%%%%%%%%%%
%%%%%%%%%%%%%%%%%%%%%%%%%%%%%%%%%%%%%%%%%%%%%%%%%%%%%%%%%%%%%%%%%%%%%%%%%%%%%%%%%%%%%%%%%%%%%%%%%%%%%%%%%%%

\begin{figure}[!ht]
    \centering
    \captionsetup{justification=justified}
    \includegraphics[width=0.8\linewidth]{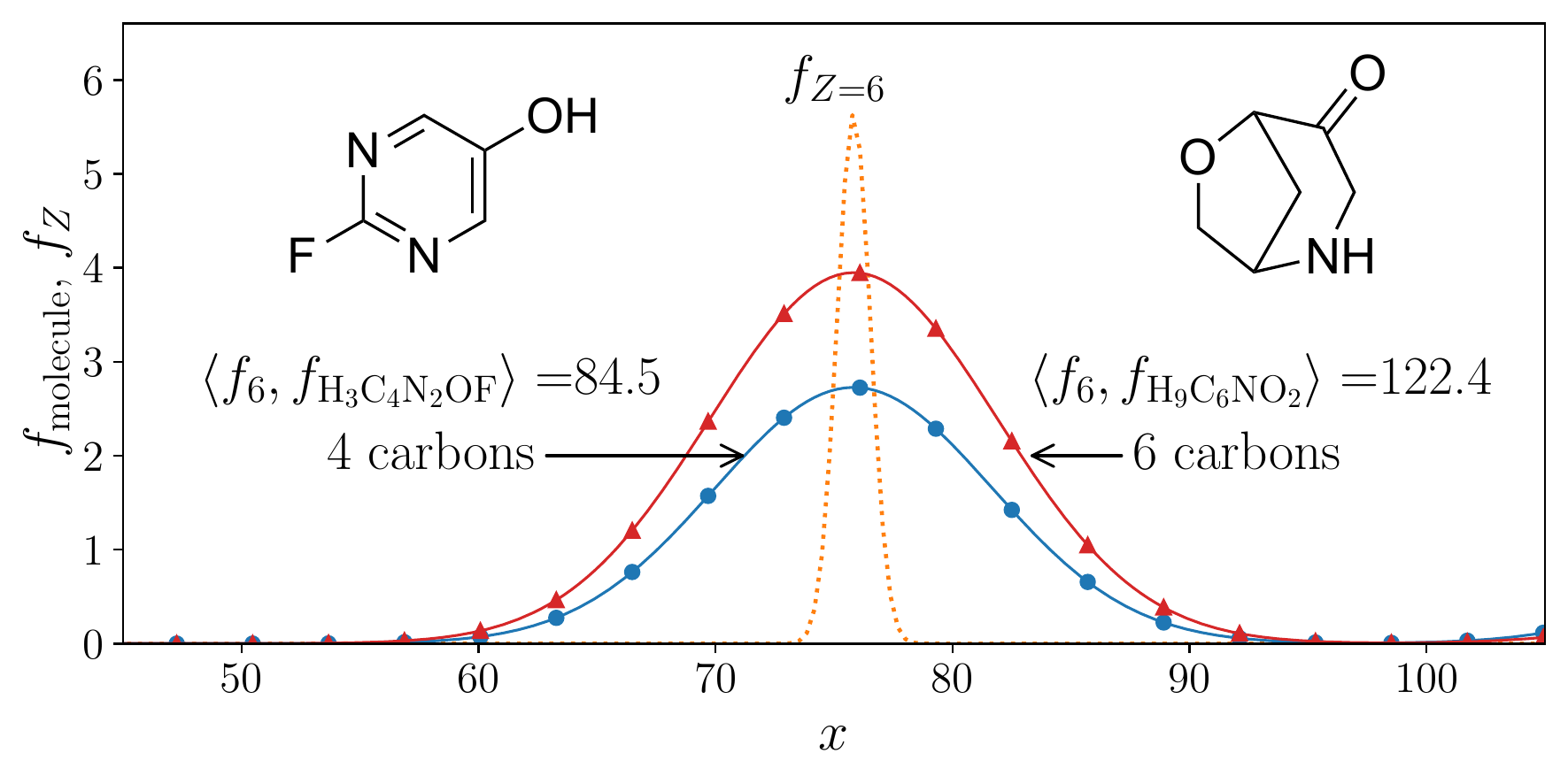}
    \caption{\justifying The carbon peaks in the molecular
    spectra of ${\rm H_{3}C_{4}N_{2}OF}$ (blue line with circles)
    and ${\rm H_{9}C_{6}NO_{2}}$ (red line with triangles).
    The values of inner products between atomic reference
    probability distribution $f_{Z=6}$ (orange dotted line)
    and each molecular function $f_{m}$ are displayed.
    All constants in~\cref{eq:N_Z} and~\ref{eq:f_molecule}
    are the same as in~\cref{fig:fig2}.
    }
    \label{fig:fig3}
\end{figure}

Since the atomic reference probability density functions are fixed,
the molecular descriptor value $\braket{f_{Z}, f_{m}}$ varies across molecules solely due to differences in the shapes of the corresponding peaks in $f_{m}$. 
For example,~\cref{fig:fig3} compares the carbon peaks of two molecular functions, $f_{\rm H_{3}C_{4}N_{2}OF}$ and $f_{\rm H_{9}C_{6}NO_{2}}$. 
Because ${\rm H_{9}C_{6}NO_{2}}$ contains more carbon atoms, the resulting carbon peak is higher and broader, yielding a larger molecular descriptor value: $\braket{f_6, f_{\rm H_{9}C_{6}NO_{2}}} = 122.4$ compared with $\braket{f_6, f_{\rm H_{3}C_{4}N_{2}OF}} = 84.5$.

To capture the peak characteristics of $f_{m}$, which are largely governed by the number and type of atoms, we introduce a model that assigns an estimated probability distribution $\hat{f}_{\nu, Z}$ to each combination of $\nu$ atoms and nuclear charge $Z$.
This formulation establishes a direct mapping between the molecular descriptors $\braket{f_{Z}, f_{m}}$, which encode peak-shape information, and the underlying atomic composition.
The estimated distributions are defined as
\begin{equation}\label{eq:f_nz_estimate}
    \hat{f}_{\nu, Z} = \sum_{i}^{\nu} \omega\times f_i\left[x \mid \mu = \left(\frac{Z^{2.4}}{2}\right)^{\beta}, \sigma^2 = \left(\theta\cdot\hat{M}_{ij}(Z, \nu)\right)^{s}\right],
\end{equation}
where $\theta$ is a parameter for adjusting the variance, and the constants $\omega$, $\beta$, and $s$ are global parameters shared with Eq.~(\ref{eq:N_Z}, \ref{eq:f_molecule}) to ensure consistency across all peak definitions.
The function $\hat{M}_{ij}(\cdot)$ approximates the average value off-diagonal elements of the Coulomb matrix, which is dependent on $Z$ and $\nu$ .
The specific parameter setting and their physical justification are discussed in Sec.~\ref{subsec:result-inverse-mapping}, where we analyze the accuracy of the inverse mapping.
%%%%%%%%%%%%%%%%%%%%%%%%%%%%%%%%%%%%%%%%%%%%%%%%%%%%%%%%%%%%%%%%%%%%%%%%%%%%%%%%%%%%%%%%%%%%%%%%%%%%%%%%%%%
%%%%%%%%%%%%%%%%%%%%%%%%%%%%%%%%%%%%%%%%%%%%%%%%%%%%%%%%%%%%%%%%%%%%%%%%%%%%%%%%%%%%%%%%%%%%%%%%%%%%%%%%%%%

\section{Results}\label{sec:results}
\subsection{Parameterization and Analysis of Inverse Mapping}\label{subsec:result-inverse-mapping}
A key step in applying the inverse mapping model is the specification of parameters that govern the estimated probability distributions $\hat{f}_{\nu, Z}$ in~\cref{eq:f_nz_estimate}.
We define the estimation function $\hat{M}_{ij}(Z, \nu)$ as
\begin{equation}\label{eq:hat_Mij}
    \hat{M}_{ij}(Z, \nu) = \frac{Z\cdot Z_{\mu}}{\rho_{Z}(1+k\nu)},
\end{equation}
where $Z_{\mu}$ is a nuclear-charge constant, $\rho_Z$ is a constant related to atom separation, and $k$ is a scaling parameter that determines how strongly $\nu$ influences the denominator.
In our implementation, we set $Z_{\mu} = 6.2$, corresponding to the average nuclear charge of H, C, N, O, and F, which are the elements present in the QM9 dataset.
For distributions estimating heavy-atoms peaks ($Z>1$), we set $k = 0.01$.
For $\hat{f}_{\nu, Z=1}$, we adopt a large value of $k = 0.2$, which yields more accurate estimates when compared with the hydrogen peaks of $f_{m}$ from QM9.
We attribute the higher $k$ value for hydrogens to two factors: (i) hydrogen atoms typically occupy terminal positions in molecules; and (ii) an increasing number of hydrogen implies a greater proportion of single bonds and, consequently leading to more elongated molecular geometries. 
Both effects contribute to a more rapid increase in the average separation between hydrogen and other atoms compared with that observed for heavier elements.

Table~\ref{table:rho_Z_vals} lists the $\rho_Z$ values used for each atom type, which produce a set of $\hat{f}_{\nu, Z}$ capable of estimating the observed peak shapes.
%%%%%%%%%%%%%%%%%%%%%%%%%%%%%%%%%%%%%%%%%%%%%%%%%%%%%%%%%%%%%%%%%%%%%%%%%%%%%%%%%%%%%%%%%%%%%%%%%%%%%%%%%%%
\begin{table}[!hp]
\begin{center}
\captionsetup{justification=justified}
  \begin{tabular}{p{2cm}p{2cm}p{2cm}p{7cm}}%{|p{2cm}|p{2cm}|p{2cm}|p{5cm}|p{2cm}|}%{  l  l  l  l  l  }
    \hline % \hline
    Atom & $Z $ & $\rho_Z$~(\AA) & Reference bond(s) (bond length~(\AA))\\ \hline  \hline
    H & 1 & 1.09 & C-H (1.07)\\ \hline
    C & 6 & 2 & C-C (1.54)\\ \hline
    N & 7 & 1.43 & C-N (1.43)\\ \hline
    O & 8 & 1.4 & C-O (1.43) and C=O (1.21) \\ \hline
    F & 9 & 1.35 & C-F (1.35) \\ \hline
    \hline %\hline
  \end{tabular}
\caption{\justifying Values of the atom separation constant ($\rho_Z$) in~\cref{eq:f_nz_estimate} for all atom types present in the QM9 dataset.} 
\label{table:rho_Z_vals}
\end{center}
\end{table}
%%%%%%%%%%%%%%%%%%%%%%%%%%%%%%%%%%%%%%%%%%%%%%%%%%%%%%%%%%%%%%%%%%%%%%%%%%%%%%%%%%%%%%%%%%%%%%%%%%%%%%%%%%%
The last column of the table lists reference bond lengths for each atom type.
Except for carbon, most $\rho_Z$ values closely match these references.
Because organic molecules are typically built on carbon skeletons, carbon atoms are more likely to be located at the molecular periphery.
This explains why the estimated average separation for carbon $\rho_{Z=6}$ is greater than even the long ${\rm C-C}$ single bond.
Multiplying $\hat{M}_{ij}$ by $\theta$ in~\cref{eq:f_nz_estimate} parallels summing the $M_{ij}$ terms in~\cref{eq:f_molecule}.
Table~\ref{table:theta_vals} lists the tested $\theta$ values that yield $\hat{f}_{\nu, Z}$ to accurately reproduce atomic peaks in $f_{m}$ for the QM9 dataset. Optimal $\theta$ values depend on both $Z$ and $\nu$.
%%%%%%%%%%%%%%%%%%%%%%%%%%%%%%%%%%%%%%%%%%%%%%%%%%%%%%%%%%%%%%%%%%%%%%%%%%%%%%%%%%%%%%%%%%%%%%%%%%%%%%%%%%%
\begin{table}[!h]
\begin{center}
\captionsetup{justification=justified}
  \begin{tabular}{p{5cm}p{4cm}p{4cm}}%{|p{2cm}|p{2cm}|p{2cm}|p{5cm}|p{2cm}|}%{  l  l  l  l  l  }
    \hline % \hline
    Atoms ($Z$) & Condition & $\theta$ value\\ \hline  \hline
    H (1); F (9) & $\nu < 4$ & $\nu +2$\\ \hline
    H (1); F (9) & $\nu \geq 4$ & $\nu$\\ \hline
    C (6); N (7); O (8) & $\nu < 6$ & $7$\\ \hline
    C (6); N (7); O (8) & $\nu \geq 6$ & $\nu$\\ \hline
    \hline %\hline
  \end{tabular}
\caption{\justifying Values of the scaling constant $\theta$ in~\cref{eq:hat_Mij} depending on $\nu$ and $Z$.} 
\label{table:theta_vals}
\end{center}
\end{table}
%%%%%%%%%%%%%%%%%%%%%%%%%%%%%%%%%%%%%%%%%%%%%%%%%%%%%%%%%%%%%%%%%%%%%%%%%%%%%%%%%%%%%%%%%%%%%%%%%%%%%%%%%%%

\begin{figure}[!h]
    \centering
    \captionsetup{justification=justified}
    \includegraphics[width=.9\linewidth]{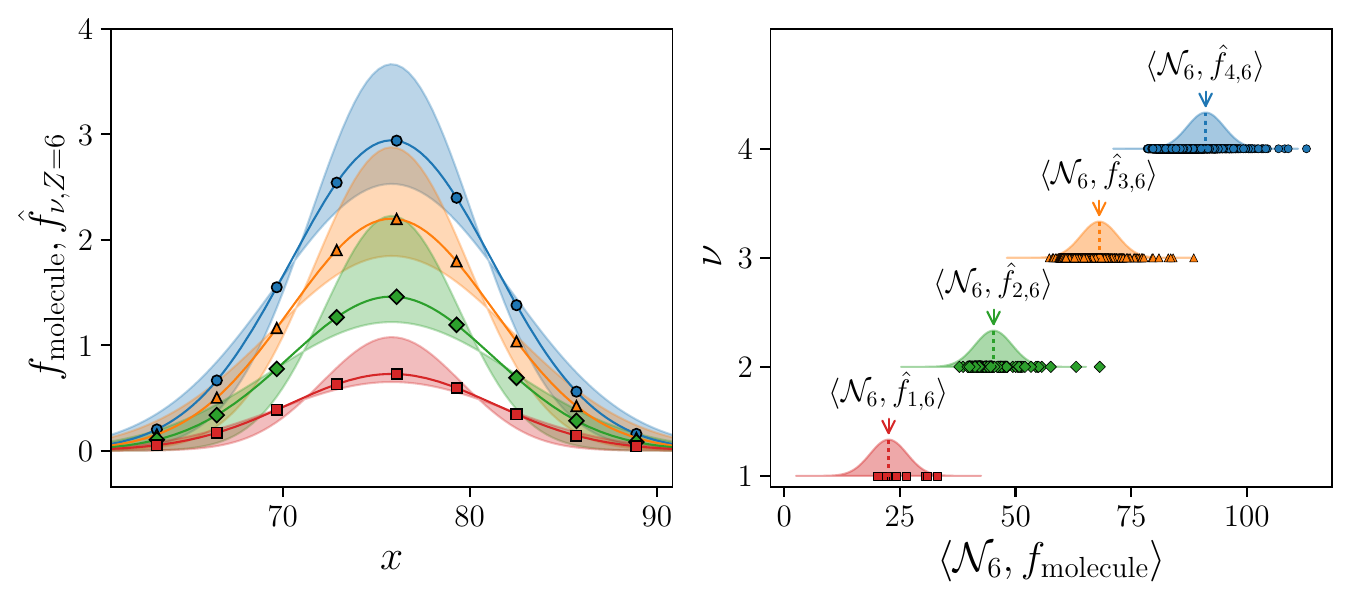}
    \caption{\justifying Left: Estimated probability distributions $\hat{f}_{\nu, Z=6}$ for $\nu$ carbon atoms.
    The shaded regions indicate the range of $f_{\rm molecules}$ corresponding to each $\nu$.
    Results are shown for $\nu = 1$ (red line with squares), $\nu = 2$ (green line with diamonds), $\nu = 3$ (orange line with triangles), and $\nu = 4$ (blue line with circles).
    Right: Distributions of $\braket{f_6, f_{m}}$ for molecules with various $\nu$ values.
    For each $\nu$, a Gaussian distribution $h_{\nu, Z}$ with the mean $\mu = \braket{f_{6}, \hat{f}_{\nu, Z=6}}$ and standard deviation $\sigma = 4$ is overlaid.
    }
    \label{fig:fig4}
\end{figure}

Fig.~\ref{fig:fig4} presents examples of the estimated atomic peaks $\hat{f}_{\nu, Z}$.
The left panel shows the estimated distributions $\hat{f}_{\nu, Z=6}$ for $1\leq \nu \leq 4$ carbon atoms.
The red line with squares corresponds to $\nu = 1$, green line with diamonds to $\nu = 2$, orange line with triangles to $\nu = 3$,
and blue line with circles to $\nu = 4$.
Each curve is overlaid with a shaded region of the same color, representing the range of $f_{m}$ observed in the QM9 molecules with the corresponding $\nu$.
All $\hat{f}_{\nu, Z}$ lie within their respective shaded ranges, indicating that the estimated distributions capture the observed $f_{m}$ carbon peak features across different $\nu$.
The right panel shows the distributions of molecular descriptor $\braket{f_{6}, \hat{f}_{\rm molecule}}$ for QM9 molecules with varying $\nu$.
For each $\nu$, a Gaussian distribution
\begin{equation}\label{eq:h_nu,Z}
h_{Z, \nu} = f(x\mid \mu=\braket{f_{6}, \hat{f}_{\nu, Z=6}}, \sigma = 4)    
\end{equation}
overlaps the scattered points along each $\nu$.
The close agreement between the Gaussian curves and the observed data confirms that $\braket{f_{6}, \hat{f}_{\nu, Z=6}}$ provides an accurate estimate of $\nu$ from the molecular descriptor.
%%%%%%%%%%%%%%%%%%%%%%%%%%%%%%%%%%%%%%%%%%%%%%%%%%%%%%%%%%%%%%%%%%%%%%%%%%%%%%%%%%%%%%%%%%%%%%%%%%%%%%%%%%%
%%%%%%%%%%%%%%%%%%%%%%%%%%%%%%%%%%%%%%%%%%%%%%%%%%%%%%%%%%%%%%%%%%%%%%%%%%%%%%%%%%%%%%%%%%%%%%%%%%%%%%%%%%%

\begin{figure}[!ht]
    \centering
    \captionsetup{justification=justified}
    \includegraphics[width=0.95\linewidth]{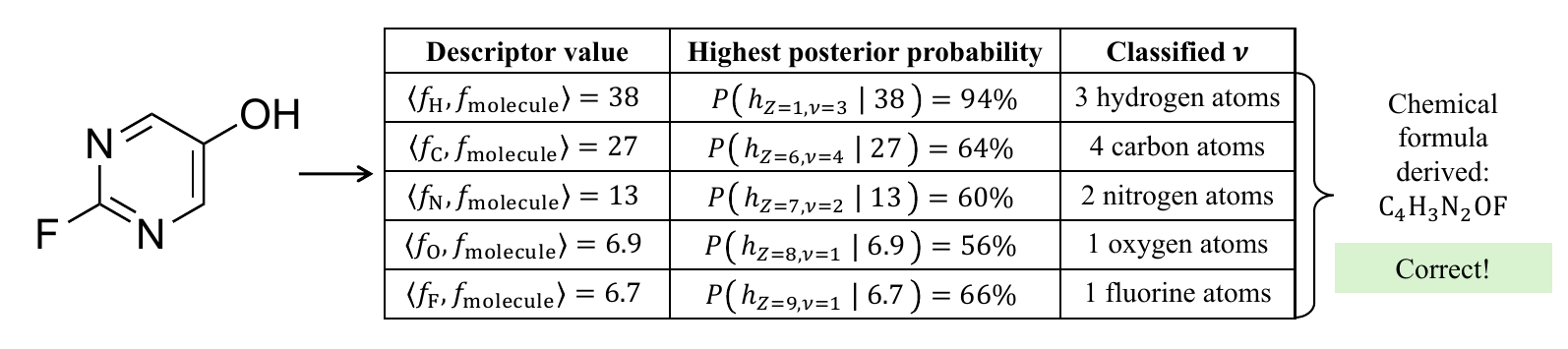}
    \caption{\justifying An example application of the inverse chemical formula mapping algorithm with optimized hyperparameters for determining the stoichiometric coefficients from a set of molecular descriptors.
    % using ${\rm H_{3}C_{4}N_{2}OF}$ as an example.
    }
    \label{fig:fig5}
\end{figure}

Using the Bayes' theorem, we classify $\nu$ from the molecular descriptor.
For a given descriptor value $f_{Z, m} = \braket{f_{Z},f_{m}}$, the probability that it originates from the distribution $h_{Z, \nu}$ is
\begin{equation}
    P(h_{Z, \nu}\mid f_{Z, m}) = \frac{h_{Z, \nu}(f_{Z, m})}{\sum_{\nu'}^{\nu_{\rm max}}h_{\nu',Z}(f_{Z, m})},
\end{equation}
where $h_{Z, \nu}(f_{Z, m})$ denotes the likelihood of $f_{Z, m}$ under the Gaussian distribution parameterized by $(\nu, Z)$. Assuming equal priors for all $h_{Z, \nu}$, the denominator normalizes the probabilities.
Fig.~\ref{fig:fig5} shows an example of how this classification algorithm successfully maps molecular descriptors to the corresponding chemical formula. 

\subsection{Molecular property optimization}
We evaluate the performance of the optimization algorithm by examining how quickly it converges to molecular structures with the desired chemical properties.
Fig.~\ref{fig:fig7} presents the optimization results for molecular entropy ($S\times T$, top panel) and zero-point vibrational energy (ZPVE, bottom panel).
Target values are uniformly sampled across the property range, with molecules of varying sizes drawn from each interval.
For entropy, molecules are sampled at intervals of $1\,{\rm kcal\,mol^{-1}}$, while for ZPVE the interval is $5\,{\rm kcal\,mol^{-1}}$.
For each target value, ten independent BO runs are performed, where the initial training datasets $\mathcal{D}_0$ are different due to randomization of the LHS method.
The bars indicate the average number of BO iterations required to reach the targets, and the error bars show the full range of the number of iterations required for successful optimization, from minimum to maximum, across all trials.
Solid green bars correspond to $100\%$ success rates, where all trials converge to the optimal structure, while open bars indicated $<100\%$ success, with the corresponding success rate labeled above each bar.
Among the partially successful cases, results with $\geq 90\%$ success are shown in green, and those with lower rates are in orange.
The gray-shaded region marks a property interval where no successful optimizations are achieved.
%%%%%%%%%%%%%%%%%%%%%%%%%%%%%%%%%%%%%%%%%%%%%%%%%%%%%%%%%%%%%%%%%%%%%%%%%%%%%%%%%%%%%%%%%%%%%%%%%%%%%%%%%%%
\begin{figure}[!ht]
    \centering
    \begin{subfigure}{.9\linewidth}
        \includegraphics[width=1\linewidth]{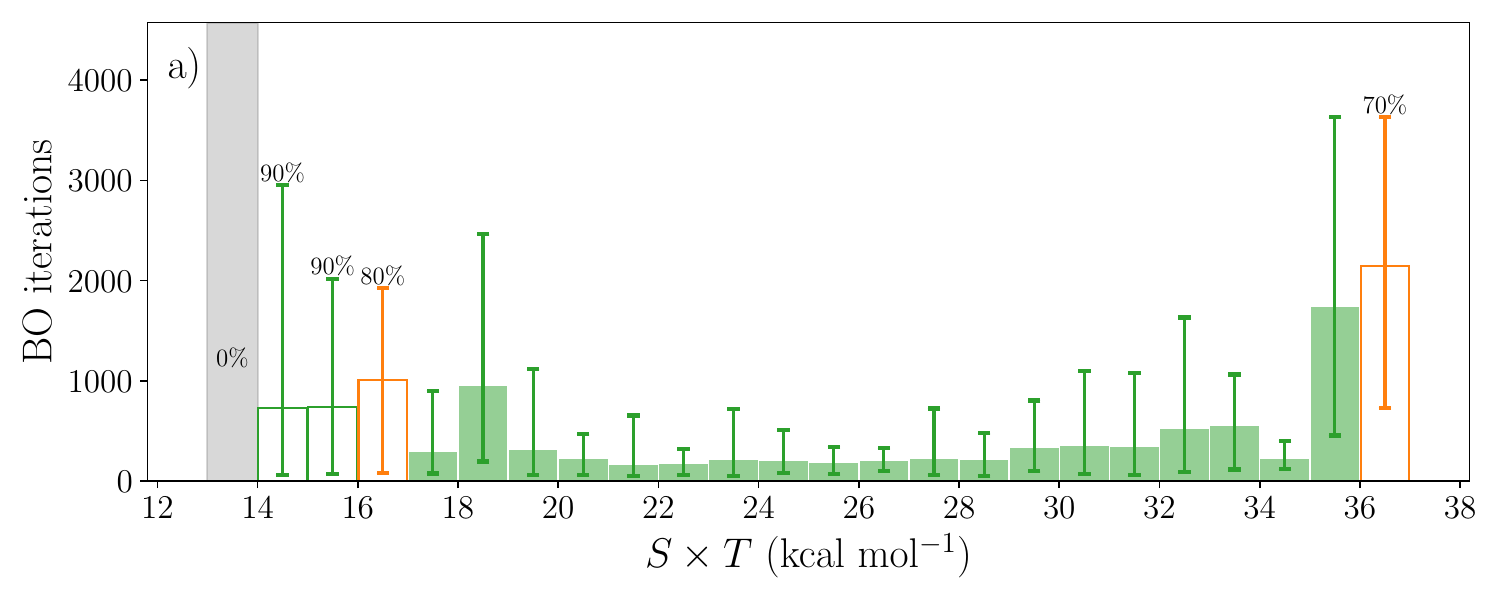}
    \end{subfigure} 
    \begin{subfigure}{.9\linewidth}
        \includegraphics[width=1\linewidth]{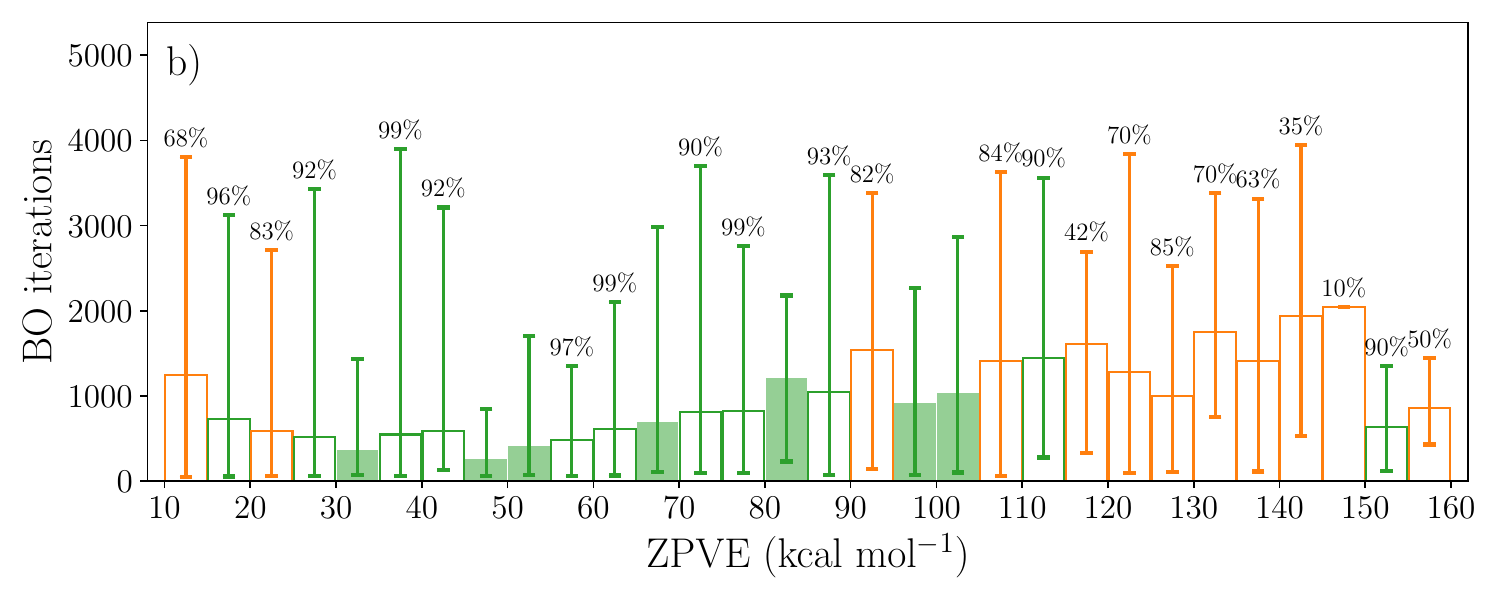}
    \end{subfigure} 
    \captionsetup{justification=justified}
    \caption{\justifying Molecular optimization results for entropy (top) and ZPVE (bottom).
    Target values are sampled uniformly across the full property range.
    For each target value, ten independent optimizations are performed.
    The bars represent the average number of BO iterations required to identify the target, and the error bars show the full range of the number of iterations across all trials. The solid green bars indicate 100$\%$ success rate, while the open bars correspond to cases with $<100\%$ success rate. The gray shaded region marks the $0\%$ success rate. For all calculations, the threshold parameter $\epsilon = 0.1\,{\rm kcal\,mol^{-1}}$.
    }
    \label{fig:fig6}
\end{figure}
%%%%%%%%%%%%%%%%%%%%%%%%%%%%%%%%%%%%%%%%%%%%%%%%%%%%%%%%%%%%%%%%%%%%%%%%%%%%%%%%%%%%%%%%%%%%%%%%%%%%%%%%%%%
The entropy values in the QM9 dataset span the range 13 to 37 ${\rm kcal\,mol^{-1}}$. As shown in~\cref{fig:fig6}(a), our algorithm achieves a 100$\%$ success rate in identifying molecules when the target entropy lies within $17\,{\rm kcal\,mol^{-1}} \leq S\times T \leq 36\,{\rm kcal\,mol^{-1}}$, typically requiring fewer than $1000$ BO iterations on average. 
Near the boundaries of the distribution, however, the optimization becomes less reliable: the success rate drops below 100 $\%$ and the required number of iterations increases.

In particular, the algorithm fails completely in the low-entropy regime ($13\,{\rm kcal\, mol^{-1}} \leq S\times T \leq 14\,{\rm kcal\, mol^{-1}}$).
The target molecule in this range that the algorithm cannot identify is water (${\rm H_2O}$).
Further analysis in~\cref{fig:fig7}(a) reveals that the success rates (blue circles) are abnormally low for molecules with only one heavy atom.
In contrast, the success rate for molecules with at least two heavy atoms remains above 90$\%$
% water (${\rm H_2O}$), ammonia (${\rm NH_3}$) and methane (${\rm CH_4}$) in the QM9 dataset.

 %%%%%%%%%%%%%%%%%%%%%%%%%%%%%%%%%%%%%%%%%%%%%%%%%%%%%%%%%%%%%%%%%%%%%%%%%%%%%%%%%%%%%%%%%%%%%%%%%%%%%%%%%%%
\begin{figure}[H]
    \centering
    \begin{subfigure}{.49\linewidth}
        \includegraphics[width=1\linewidth]{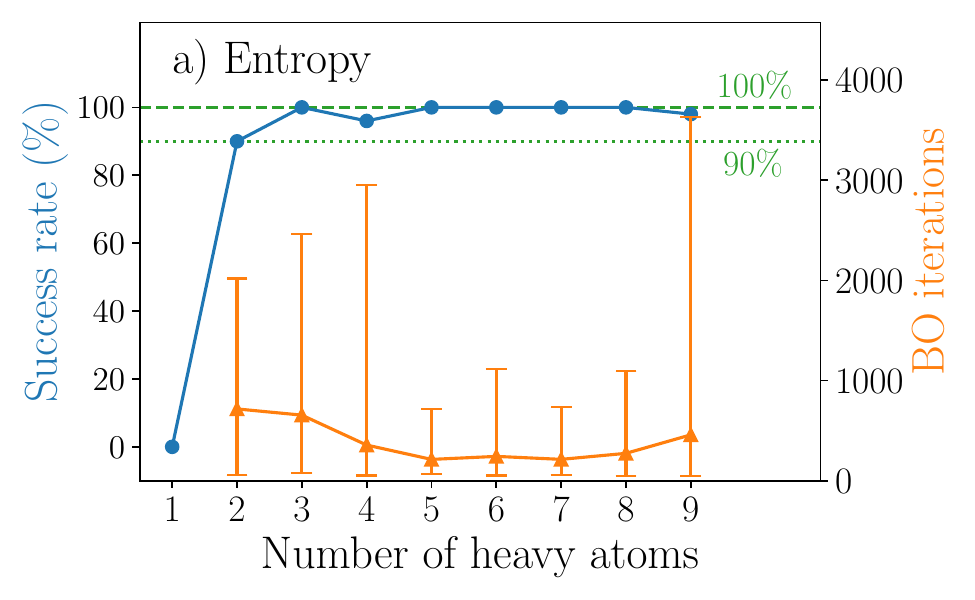}
    \end{subfigure} 
    \begin{subfigure}{.49\linewidth}
        \includegraphics[width=1\linewidth]{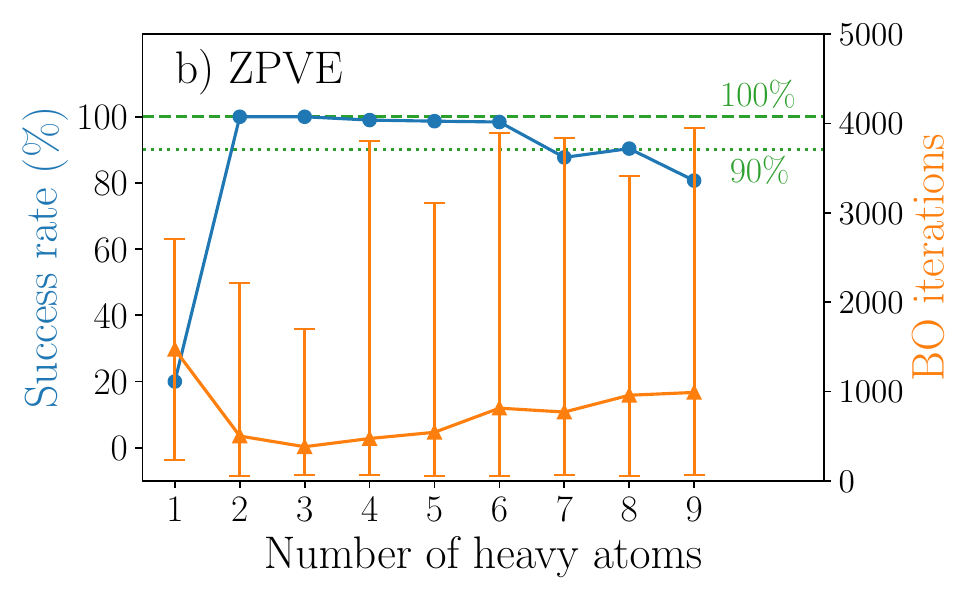}
    \end{subfigure} 
    \captionsetup{justification=justified}
    \caption{\justifying Correlation between the number of heavy atoms with success rate (blue circles) and with BO iterations (orange triangles). Left panel: entropy optimization. Right panel: ZPVE optimization.
    }
    \label{fig:fig7}
\end{figure}
%%%%%%%%%%%%%%%%%%%%%%%%%%%%%%%%%%%%%%%%%%%%%%%%%%%%%%%%%%%%%%%%%%%%%%%%%%%%%%%%%%%%%%%%%%%%%%%%%%%%%%%%%%%

For ZPVE, the values in the QM9 dataset span a much broader range, from 10 to 160 ${\rm kcal\,mol^{-1}}$.
As shown in~\cref{fig:fig6}(b), optimizing molecular structures with respect to ZPVE is more challenging than for entropy.
Overall, the required number of BO iterations is significantly higher, and only seven intervals achieve $100\%$ success rate.
Success rates vary substantially across the ZPVE range, with values spanning from as high as $99\%$ to as low as $10\%$.
Unlike the entropy case, where failures are clearly concentrated at the extreme ends of the distribution in~\cref{fig:fig6}(a), ZPVE failures are scattered across the property range.
This indicates that ZPVE optimization is more sensitive to local structural features of molecules and is not solely governed by the extremity of the target value.
Fig.~\ref{fig:fig7}(b) shows that molecules with 2 $-$ 6 heavy atoms achieve nearly $100\%$ success rates with fewer than 1000 BO iterations on average.
However, as the number of heavy atoms increases, the success rate declines while  the average number of iterations grows. 
The algorithm performs the worst for molecules with only one heavy atom, similar to the entropy result.
For these molecules, the overall success rate drops to just $20\%$, and convergence requires the highest average number of iterations ($\approx 1500$ iterations). 

\section{Summary}
In this work, we present a data-efficient framework for molecular property optimization in discrete chemical compound spaces based on Bayesian optimization (BO).
Our approach addresses the fundamental challenges that limit molecular optimization:
the discreteness and high dimensionality of chemical compound space,
as well as the inverse problem of mapping numerical molecular descriptor vectors
back to chemically valid structures.
Our approach overcomes these problems through an integrated strategy that
combines three key components:
(i) a compact and physics-informed nine-dimensional molecular descriptor vector
that preserves essential chemical information
while mitigating the curse of dimensionality;
(ii) BO with Gaussian process regression,
which allows sample-efficient exploration of the descriptor landscape;
and (iii) a robust, generalizable, and chemically consistent inverse mapping scheme
that reconstructs stoichiometry and valid molecular structures
from optimized descriptor vectors.
Together, these components establish a coherent framework that enables
data-efficient optimization in a vast and discrete chemical compound space.

We benchmark the proposed framework on the QM9 dataset by optimizing molecular entropy and zero-point vibrational energy (ZPVE). For both properties, the approach achieves success rates exceeding $80\%$ for molecules containing at least two heavy atoms, typically converging to optimal structures with fewer than 1,000 Bayesian optimization iterations.
Entropy optimization is particularly effective, reaching $100\%$ success
rate while requiring fewer than 1,000 molecular evaluations in over
$80\%$ of test cases, demonstrating high sample efficiency.
Optimization performance declines near the extremes of the property distribution,
most notably for low-entropy molecules such as water (${\rm H_2O}$),
highlighting the difficulty of interpolation in sparsely sampled
regions of descriptor space.
ZPVE optimization is comparatively more challenging, with success rates more sensitive to molecular size and structural complexity.
Nevertheless, our approach consistently and efficiently navigates
chemically constrained property landscapes and exhibit reliable convergence across diverse molecular systems.

By combining chemically meaningful descriptors, Bayesian optimization,
and a physically interpretable and consistent inverse mapping scheme,
this work establishes a general strategy
for high-precision inverse molecular optimization
in discrete chemical compound space.
Beyond the specific applications studied here,
the results illustrate the importance of reducing molecular representations to
compact and physically grounded descriptors to overcome the dimensionality barrier that limits data-efficient optimization in chemistry.
The proposed framework therefore makes BO practically viable
for molecular design under small-data regimes,
where conventional generative or gradient-based approaches struggle.
Despite being demonstrated on the QM9 dataset,
our approach is not dataset-specific and can be generalized to larger chemical compound spaces,
alternative descriptor sets,
or other molecular databases.
Future developments may incorporate de novo molecular generative models~\cite{wang_cmolgpt_2023, shen_deep_2021}
or large language models (LLMs)~\cite{boiko_autonomous_2023, ruan_automatic_2024, m_bran_augmenting_2024, caldasramos_llm_review_2025, zou_agente_2025}
for structure creation instead of database search, as well as integrate uncertainty-aware or active-learning strategies to further enhance efficiency. 
In summary, the proposed framework bridges continuous optimization in descriptor space with discrete molecular design, offering a promising route to accelerate the discovery of molecules with tailored properties for applications across chemistry, material sciences, and beyond.
\section{Data availability statement}
All data that support the findings of this study are included within the article (and any supplementary files).
\section*{Acknowledgments}
This work was supported by NSERC of Canada. 
\bibliographystyle{unsrtnat} % Bibliography ordered by appearance% \bibliographystyle{iopart-num}
\bibliography{references}  %%% Uncomment this line and comment out the ``thebibliography'' section below to use the external .bib file (using bibtex) .

\end{document}